
\magnification 1200
\tolerance 2000
\font\titolo=cmbx10 scaled\magstep0

\font\myit=cmti7 scaled\magstep1

\parskip 3truemm
\hsize=140 truemm
\vsize=210 truemm
\hoffset=10 truemm

\def\.{\cdot}

\def\sn{\smallskip \noindent}

\def\mn{\medskip \noindent}

\def\bn{\bigskip \noindent}

\def\ni{\noindent}

\def\nostrocostrutto#1\over#2{\mathrel{\mathop{\kern 0pt \rlap
  {\raise.2ex\hbox{$#1$}}}
  \lower.9ex\hbox{\kern-.190em $#2$}}}
\def\lsim{\nostrocostrutto < \over \sim}
\def\gsim{\nostrocostrutto > \over \sim}

\def\bino{\tilde B}
\def\w3{\tilde W_3}
\def\hu{\tilde H_u}

\def\hd{\tilde H_d}

\def\c{\chi}

\null
\nopagenumbers
\rightline{LNGS  94/90}
\rightline{DFTT \ 5/94}
\rightline{January 1994}
\sn
\centerline{\titolo High Energy Gamma--Radiation from the Galactic }
\centerline{\titolo Center due to Neutralino Annihilation}
\mn
\centerline{V. Berezinsky$^{(a)}$, A. Bottino$^{(b)}$, G. Mignola$^{(b)}$}
\sn
\centerline{\myit {$^{(a)}$ INFN, Laboratori Nazionali del Gran Sasso,}}
\centerline{\myit {67010 Assergi (AQ), Italy and}}
\centerline{\myit {Institute for Nuclear Research, Moscow, Russia.}}
\sn
\centerline{\myit {$^{(b)}$ Dipartimento di Fisica Teorica
dell'Universit\`a di Torino}}
\centerline{\myit {and INFN, Sezione di Torino, Italy}}
\centerline{\myit {Via P. Giuria 1, 10125 Torino, Italy}}
\vskip 4truemm
\bn
\centerline{\bf Abstract}
\sn
\noindent
We study the NGS (Non--dissipative Gravitational Singularity) model, which
successfully describes the non--linear stage of evolution of perturbations (see
[1], [2] and references therein). This model predicts DM density distribution
$\rho(r) \sim r^{-\alpha}$ with $\alpha \simeq 1.8$ which holds from
very small distances $r_{\rm min} \simeq 0.01~{\rm pc}$
up to very large distances
$r_{\rm max} \simeq 5~{\rm Mpc}$. Assuming the neutralino to be a
CDM particle, we
calculate the annihilation of neutralinos in the vicinity of the singularity
(Galactic Center). If neutralinos are the dominant component of DM in our
Galaxy,
the produced energy is enough to provide the whole observed
activity of the GC. Neutralinos of the most general composition and of
mass in
the range $20~{\rm GeV} \leq m_\c \leq 1~{\rm TeV}$ are considered. We find the
 neutralino compositions which give the relic density  needed
for the Mixed Dark Matter (MDM) model and we evaluate for these compositions
the high--energy
($E_{\gamma} > 100 ~{\rm MeV}$) gamma--ray flux under the constraint that
 the radio flux is lower
than the observational limit.  The compositions with the detectable gamma--ray
flux which we found are provided by a set of
almost pure gaugino states with the neutralino mass between $100$ and $500$
GeV.

    We demonstrate that a detectable high--energy gamma--ray flux is produced
by the neutralino annihilation also in the case when neutralinos provide a
small fraction  (down to $0.1 \%$) of the DM in our Galaxy. The predicted flux
is $F_\gamma \sim 10^{-7}-10^{-8}~{\rm cm}^{-2}~{\rm s}^{-1}$ for  $E_\gamma
\gsim 300~{\rm MeV}$

\vfill
\eject

\footline{\hfill}
\headline{\hfil-- \folio\ --\hfil}
\pageno=1

1. Introduction.

   The spectrum of perturbations as observed by COBE, IRAS and CfA is most
naturally explained by the Mixed Dark Matter (MDM) model [3] with $\sim 60-70
\%$ of Cold Dark Matter (CDM), $\sim 20-30 \%$ of Hot Dark Matter (HDM) and
$\sim 10 \%$ of baryonic matter . The most plausible candidates for the MDM
particles are the $\tau$--neutrino as a HDM particle
and the neutralino $\c$ as a CDM
particle. According to this model $\delta \rho / \rho$ on the galactic scale
$\lambda \sim 1~{\rm Mpc}$ is dominated by the CDM particles, while on the
scales  $\lambda \sim 10-100~{\rm Mpc}$ $\delta \rho / \rho$ is dominated by
the HDM particles. The baryons, which decoupled from the radiation much later
than
CDM particles, streamed into the potential wells built by the space
distribution of CDM particles. As a rough estimate of the average fraction of
the baryon density in the halo one can take the relative density of baryons in
the whole Universe ($\sim 10\%$) and this assumption does not contradict the
recent observations of MACHO candidates[4].

    When $\delta \rho / \rho$ on the galactic scale
approaches one, the evolution
of fluctuations enters the non--linear stage. There are two solutions for this
stage, determined by the space symmetry of the initial fluctuations, i.e. of
those with $\delta \rho / \rho \sim 1$. The first solution was found by
Zeldovich [5] for a case of asymmetric initial
distribution of matter relative to the
local maximum of density. In this case the evolution terminates with the
production of the flat formations -- the Zeldovich pancakes. For this solution
the potential well within a galaxy is not formed and the CDM gas is not
self--trapped.

    The other solution, found first in ref. [1], exists if the initial
distribution is more symmetric. In this case a strong singularity is formed and
the CDM gas is self--trapped. Each singularity, which is called
Non--dissipative Gravitational Singularity (NGS), is associated with a galaxy.
The main predictions of this model according to ref. [2], can be summarized as
follows: \hfill \break
\indent i) The density distribution of CDM particles centered at a NGS is
given by
$$\rho (r) \propto r^{-\alpha} \eqno(1)$$
with $\alpha \simeq 1.8$. The distribution (1) is
valid down to very small distances $r_{\rm min}$ where a cut off
can be caused by several phenomena[2]. The largest value of
$r_{\rm min}$ is related to the case of a black hole
in the NGS. For a black hole with a mass $M \sim 10^{6}~M_{\odot}$ this value
is $r_{\rm min} \sim 0.01~{\rm pc}$.
The maximum value up to which the
distribution (1) holds is not predicted by the NGS model, because this is the
initial condition for the non--linear problem. We suggest that the value of
$r_{\rm max}
\sim 5~{\rm Mpc}$, as it is given by the two--point correlation function, is
determined by the scale where HDM dominates in the perturbation amplitudes. The
distribution (1) is in excellent agreement with the observations. \hfill \break
\indent ii) The predicted density of CDM at the distance $r_{\odot}=8.5~{\rm
kpc}$ from the Galactic Center is equal to
$\rho^{\rm loc}_{\rm CDM} \simeq 0.2~{\rm GeV}$, if we take $<\rho_{\rm CDM}>
\approx 0.7 \, \rho_{c}$
($\rho_c = 1.88\cdot 10^{-29}h^2~{\rm g}~{\rm cm}^{-3}$). This result is
in a good agreement with the DM density obtained from the
rotational curves in our Galaxy, $0.3 ~{\rm GeV}~{\rm cm}^{-3}
\leq \rho^{\rm loc}_{\rm DM} \leq 0.43
{}~{\rm GeV}~{\rm cm}^{-3}$, for the total DM density. \hfill \break
\indent iii) Annihilation of CDM particles in the vicinity of the singularity
results in a pointlike radiation source in the Galactic Center.

    In ref.[2] the luminosity due to the neutralino annihilation in the NGS
model has been evaluated, together with the ensuing
gamma--ray and radio fluxes. Comparisons of the calculated fluxes with
the available experimental upper limits were then used in ref.[2] to
obtain lower limits for neutralino masses in the case of almost pure
neutralino configurations (either gauginos or higgsinos) and under the
hypothesis that neutralinos provide a fraction of about $70 \%$ of DM density
in the Galactic halo.
In ref.[2] the theoretical evaluations referred mainly to not too
heavy neutralinos, i.e. $m_\c \leq O(m_W)$.

    In this paper we present a more general analysis valid for a wide range of
neutralino masses ( $20~{\rm GeV} \leq m_\c \leq 1$ TeV) and of neutralino
compositions and we mainly address the question: can the annihilation of
neutralinos around the singularity result in detectable gamma--ray flux at
energy $E_\gamma \gsim 100 ~{\rm MeV}$? Apart from the favourite case when the
neutralino is the only CDM particle, we shall study the case of arbitrary
$\Omega_\c = \rho_\c / \rho_c$, which in the  framework of the MDM model
implies
the presence of other CDM particles X with $\Omega_X > \Omega_\c$

2. Neutralino

        We shall consider the neutralino in the context of the
Minimal Supersymmetric Standard Model (MSSM). The neutralino
is defined as the lowest--mass linear combination

$$ \c = Z_{11} \w3 + Z_{12} \bino+ Z_{13} \hu+ Z_{14} \hd \eqno(2)$$

\noindent where $ \tilde W_3$ and $\tilde B$ are the SU(2) and U(1)
 neutral gauginos and $ \tilde H_{u,d}$ are
the higgsinos. The space of neutralino states is determined by three
independent parameters: the masses of $\bino$ and $\w3$, $M_1$ and $M_2$,
respectively, with the  GUT relation between them $M_1 \simeq 0.5~ M_2$,
the Higgs mixing parameter $\mu$, and the ratio of the two
v.e.v.'s which give masses to up--type and down--type quarks,
$\tan \beta = v_u/v_d$.
The natural range for $\tan\beta$ is: $1 \leq \tan\beta \leq m_t/m_b$.
In the calculations discussed below the parameters $M_2$ and $\mu$ are
varied in the wide ranges: $0< M_2 < 6$ TeV and $-3 ~{\rm TeV} < \mu < 3~ {\rm
TeV}$.

        The neutralino is assumed to be the lightest SuSy
particle and to be stable because of R--parity conservation.
The decoupling of neutralinos in the early Universe with the subsequent
annihilation results in the neutralino relic abundance

$$\Omega_{\c} h^{2}=2.13\cdot 10^{-11}
\left(T_{\chi}\over T_{\gamma}\right)^{3}
\left(T_{\gamma}\over 2.7^\circ K\right)^{3}N_{f}^{1/2}\left(GeV^{-2}
\over ax_{f}+{1\over 2}bx^{2}_{f}\right),\eqno(3)$$

\ni where h is the dimensionless Hubble parameter,
$x_f = T_f/m_\c$, $T_f$ is the neutralino freeze--out temperature,
$T_\gamma$ and $T_\c$ are the present temperatures for relic photons and
neutralinos, $N_f$ denotes the effective number of degrees of freedom
at the freeze--out temperature.
$N_f \sim 100$ for $m_\c \geq m_W$.
The thermal average of the $\c\c$--annihilation cross--section $< \sigma v >$
at the temperature $T=x m_\c$ is parameterized in (3) as

$$< \sigma v > = a + bx \quad .\eqno(4) $$

\ni The expression  $a x_f + 1/2~ b x_f^2 \equiv < \sigma v >_{\rm int}$
in eq.(3)
represents the integrated cross section from $T_f$ to the present
temperature.

   Since we wish to discuss neutralinos with masses up to 1 TeV,
many channels have to be considered in the neutralino annihilation
processes. In the present paper, for the values of $<\sigma v>$ and of
the neutralino relic abundance, we employ the results of the thorough
analysis of ref.[6], where the full set of possible final states have
been considered: fermion--antifermion pair, pair of two (neutral and
charged) Higgses, one Higgs boson--one gauge boson states and pair of two gauge
bosons.
In the evaluation of $\sigma$, the exchanges of the following particles
: $ Z$ boson, Higgses, sfermions $\tilde f$, neutralinos and charginos, have
been included depending on the specific final state.
As for the values of the masses of the
neutral Higgs bosons, the
two CP--even: h, H (of masses $m_h$ and $m_H$, with $m_h < m_H$) and the
CP--odd: A (of mass $m_A$), we employed the usual relationships which
include radiative corrections. These imply that if,
for instance, $m_h$ is used as a
free parameter, then $m_A$ and $m_H$ turn out to be dependent on $m_h$,
$\tan \beta$ and also (through the radiative terms) on $m_t$ (top--quark mass)
and $ \tilde m$ (mass of the top scalar partners,
considered here as degenerate).
We remind that the LEP lower limit for $m_h$ is $m_h > 50~ {\rm GeV}$  [7].

In our further calculations we give most attention to those compositions of
$\c$
which provide
$\Omega_\c \simeq 0.6-0.7$. However, we shall also consider a case of
smaller $\Omega_\c$, which implies that neutralino compose only a fraction
of CDM in the Universe.

3. The pointlike source in the Galactic Center due to $\c\c$--annihilation

        For the distribution (1) annihilation between neutralino pairs would
 be very efficient in a
region of linear size $\simeq r_{min} \simeq 0.01~{\rm pc}$ which implies for
the source the
angular size  $\simeq 10^{-6}-10^{-7}$ rad. The total  luminosity
of the source  is

$$ L = 2 m_\c <\sigma v >_0 P \eqno(5)$$

\noindent where  P is

$$ \eqalignno{
P & \equiv
{1 \over m_{\c}^2} \int dr 4 \pi r^2 \rho_{\c}^2 (r) \cr
&= {{2 \pi} \over 0.3} {1 \over m_{\c}^2}
\left( \rho_\c^{\rm loc}\right)^2 r_{\odot}^3
\left ( {r_{min} \over r_\odot} \right )^{-0.6} &(6)\cr}$$

\noindent and $<\sigma v>_0$ denotes the value of $<\sigma v>$ at the present
temperature.

Here we have employed for the neutralino density profile in our Galaxy the NGS
distribution of Eq.(1) and have denoted by $\rho_\c^{\rm loc}$  the local
(solar
neighborhood) neutralino density.
 Due to different space distribution of CDM and HDM in the
Universe, we expect $\rho_\c^{\rm loc} / \rho_{\rm DM}^{\rm loc}
\gsim \Omega_\c / \Omega$. However, for
simplicity we used in our calculations $\rho_\c^{\rm loc} /
\rho_{\rm DM}^{\rm loc}
= \Omega_\c / \Omega$. For $r_{\rm min}$
 we shall use the capture radius of the black hole with
mass $M \sim 10^6M_\odot$ [2]: $r_{\rm min} \simeq 0.01~{\rm pc}$.

In the evaluation of cross section
$<\sigma v>_0$
in Eq.(5) we use the
approximation $<\sigma v>_0 \simeq a$, since at present, inside the Galaxy,
neutralinos most probably have a gaussian velocity distribution with an
average value $\bar v \simeq 300 ~{\rm km} ~{\rm s}^{-1}$ and consequently
$x \simeq \bar v^2 /6 \sim 10^{-7}$. Annihilation of neutralinos into quarks
and leptons thorough the main channels $b \bar b$, $t \bar t$, $W^+W^-$, and
$ZZ$ has been taken into account. The part of luminosity due to fermion
production is dominated by
heavy (b and t) quarks. The $b$($\bar b$) quarks, either produced directly
or by $t$($\bar t$) decays, do not decay within the confinement region, but
produce copiously lighter quarks which then turn into pions and kaons.
For the calculation of the hadronization process
we used the Lund Jetset 7.2 Monte Carlo
simulation program [8]. This program, for any given initial
configuration (i.e., the $\c-\c$ annihilation final state $F$), allows
one to follow the
hadronization and to infer the electron and $\gamma$ final distributions.
We have
performed this analysis for each final state mentioned above.
The statistical sample was taken to be $2 \cdot 10^4$ events and the relevant
physical
parameters for each process were taken from the PDG [9].
It is worth mentioning that the photon and electron spectra computed at
different center--of--mass energies $E_{\rm cm} \simeq 2 m_\c $, when plotted
in
terms of the variable $x=E/(E_{\rm cm}/2)$,
exhibit remarkable scaling properties[10].
Thus $\c\c$--annihilation in the GC results in high energy ($E_\gamma \gsim
100~{\rm MeV}$) gamma--radiation, X--rays (mainly from the Compton scattering
of the electrons), in gamma--ray lines (from $\c\c
\rightarrow \gamma\gamma$ and
$e^+e^- \rightarrow \gamma\gamma$ annihilation) and in radio--emission (from
synchrotron radiation of electrons in magnetic fields). The estimates in
ref.[2] show that expected high energy ($E_\gamma \gsim
100~{\rm MeV}$) gamma--radiation and radio--emission can  be at the level of
the
observable fluxes or may exceed them.

    The quantity of interest for observations  is the integral
gamma--ray flux

$$ F_\gamma (> E_\gamma) = { P \over { 4 \pi r_\odot^2}}
\sum_F <\sigma v>_F \int_{x_\gamma}^{\infty} dx
\left ( {{dN_\gamma (x)} \over dx} \right )_F \eqno(8)$$

\noindent where P is given by (6), $x=E_{\gamma} / m_\c$,
$<\sigma v>_F$ is the $\c\c$ annihilation cross section
in the final state $F$ and ${dN_\gamma (x)} / dx$ is the photon energy
spectrum mainly due to neutral pions decay.
The existing limit [11] on gamma--ray flux from the GC is related to
$E_\gamma \sim 300~{\rm MeV}$. More generally, for future experiments
$E_\gamma$ of interest is $\sim 70-100~{\rm MeV}$.

    The energy flux for radio--emission
at frequency $\nu$ at Earth can be calculated as

$$ F_\nu = \int^{m_\c}_0 dE_e Q_e(E_e) \int^{E_e}_{0} dE {{P(E,\nu)}
\over {b(E)}}  \eqno(9)$$

\ni where $E_e$ is the electron energy at generation, $E$ is the
electron energy at
the moment  of synchrotron radiation, $b(E)=(dE /dt)_s$ is the synchrotron
energy--loss of an electron and $P(E,\nu)$ is the energy--loss of electron with
energy $E$ due to emission of photons with frequency $\nu$. The number of
electrons with energy $E_e$, $Q(E_e)$, generated per unit time in the GC due to
$\c\c$--annihilation is given by

$$ Q(E_e)= {1 \over {m_\c}} P \sum_F <\sigma v>_F
\left ( {{dN_e (x)} \over dx} \right )_F \eqno(10)$$

\ni where $x=E_e / m_\c$ and $dN_e (x) / dx$ is defined in the same way as
$dN_\gamma (x) / dx$.

    In the calculation of radio flux we assumed the magnetic field in the GC to
be $H=10^{-2.5}~{\rm G}$ (see below).

    We shall now briefly review the
observational properties of the core of the GC following the references
[12--13]. A violent activity is observed
within a region of size $\sim 1~{\rm pc}$, which can be powered by
$\c\c$--annihilation.
The total bolometric luminosity, mostly due to the
optical and UV radiation, is $L_{\rm bol} \simeq 4 \cdot 10^{40}~{\rm erg}~{\rm
s}^{-1}$. The center of activity coincides with the compact non--thermal radio
source Sgr A$^*$. It has an unusual rising spectrum $F_\nu \simeq 1 (\nu /
10~{\rm GHz})^{1/4}~{\rm Jy}$ in the frequency range $1~{\rm GHz}
<\nu<86~{\rm GHz}$.
Its luminosity is $L_\nu = 1.3 \cdot 10^{34}~{\rm erg}~{\rm s}^{-1}$
at wavelength
$\lambda > 3~{\rm mm}$. The source is surrounded by a more extended radio
halo.

    The X--ray radiation ($0.5 - 4.5~{\rm KeV}$) from GC reveals [14] several
compact sources one of which, with luminosity $L_X \simeq
1.5 \cdot 10^{35}~{\rm
erg}~{\rm s}^{-1}$, coincides with Sgr A$^*$. The extended X--ray component
around this source has luminosity $L_X \simeq 2.2 \cdot 10^{36}~{\rm
erg}~{\rm s}^{-1}$.

    High energy gamma--radiation from GC is not detected. The COS B upper
limit at
$E_\gamma >300~{\rm MeV}$ is $F_\gamma < 4 \cdot 10^{-7}~{\rm cm}^{-2}~{\rm
s}^{-1}$.
There are indications of the presence of a powerful flux
($F_\gamma \sim 10^{-3}~{\rm cm}^{-2}~{\rm s}^{-1}$) in the form of $0.511
{}~{\rm MeV}$ e$^+$e$^-$--annihilation line from GC.
The core is submerged into a dust cloud, probably in the form of a ring with
inner and outer radii of $2$ pc and $4$ pc, respectively. The dust absorbs a
considerable part of the radiation from the core and reradiates it in the form
of IR radiation.
The phenomenological model of the core is developed in refs. [12,13].
It has been
shown that the observations provide an overdetermined set of
relations for the properties of the core; these are satisfied by a unique
 and
plausible solution [12]. In particular, the mean magnetic field in the core is
found to be $\bar H_{\perp} = 10^{-2 \pm 0.5}~{\rm G}$.

4. Results and Conclusions

    Let us turn now to the presentation and the discussion of our results.
In Fig.1 we report the values of the neutralino
relic abundance in the form of a scatter plot obtained by varying $M_2$ and
$\mu$ in the ranges $0< M_2 < 6$ TeV and $-3 ~{\rm TeV} < \mu < 3~ {\rm
TeV}$, respectively.
The other parameters are taken at
the following values: $\tan \beta = 8$, $m_h=50~{\rm GeV}$,
$m_{\tilde f} = \tilde m = 3~{\rm TeV}$ ($m_{\tilde f}$  denotes the mass of
the sfermions, taken here to be degenerate).
We notice that in this representative
point of the parameter space many neutralino configurations provide large
values for $\Omega_\c h^2$; for many configurations the
relevant $\Omega_\c h^2$ value is even above the cosmological bound
$\Omega h^2=1$. This situation occurs for gaugino--dominated compositions,
since
in this case $\c-\c$ annihilation proceeds mainly through $\tilde f$--exchange,
but this process is now strongly hindered due to the large value assigned to
$m_{\tilde f}$.

    In Fig.2 a scatter plot is displayed  for the integral gamma--ray flux
$F_\gamma(> 300~{\rm MeV})$ evaluated from eq.(8).
 We remind that in the calculation of $F_\gamma$
the rescaling for $\rho^{\rm loc}_{\c}=\Omega_\c \rho^{\rm loc}_{\rm DM}$
is used.
Fig.2 shows that a large fraction of $\c$--configurations with $m_\c
\lsim 100~{\rm GeV}$ are excluded by the COS B upper limit $F_\gamma
(> 300~{\rm MeV}) < 4 \cdot 10^{-7}~ {\rm cm}^{-2}~{\rm s}^{-1}$
[11]. For the configurations allowed by this limit, the graph does not show the
values of $\Omega_\c h^2$, i.e. it does not specify whether the neutralino can
provide $\Omega_{\rm CDM} \sim 0.7$. This informations is added in Fig.3,
where in a $\mu-M_2$ diagram are displayed the $\c$--compositions
which satisfy the two following requirements: a) $\Omega_\c$ is
within the range: $0.5 \leq \Omega_\c \leq 1$ which,
taking $h=0.7$, implies $0.25 \leq \Omega_\c h^2 \leq 0.5$; b) the relevant
radio flux, as evaluated from eq.(9), does not exceed the experimental
upper bound $F_\nu <~ 1 \cdot 10^{-23}~{\rm erg}~{\rm cm}^{-2}~{\rm s}^{-1}~
{\rm Hz}$ at $\nu=1~{\rm GHz}$ [12,13].
The neutralino configurations that meet these properties are denoted in Fig.3
by squares: they correspond to gaugino--dominated compositions with masses in
the
range $100~{\rm GeV} \lsim m_\c \lsim 500~{\rm GeV}$.
For these $\c$--configurations
the relevant gamma--ray flux is
$5 \cdot 10^{-9} ~{\rm cm}^{-2}~{\rm s}^{-1}
\lsim F_\gamma \lsim 2 \cdot 10^{-7}~{\rm cm}^{-2}~
{\rm s}^{-1}$, i.e. this source is detectable with
existing or future devices.

    Fig. 4 displays the scatter plot of $F_\gamma$ vs $\Omega_\c$ restricted to
those configurations with mass $20~{\rm GeV} \leq m_\c \leq 100~{\rm GeV}$,
which satisfy
the observational upper bound on radio--emission. The values $\Omega_\c < 0.1$
imply that CDM is composed mostly by some other particle(s), with neutralinos
providing a small part of the CDM density. Fig. 4 shows that in this case the
gamma--ray flux is marginally detectable down to
$\Omega_\c \simeq 3 \cdot 10^{-4}$.
It means that the NGS model can be tested with the help of $\c\c$--annihilation
even if the neutralino constitutes only a small part of CDM.

    The general features of the previous discussion remain valid when
we assign to the free mass parameters, $m_{\tilde f}$ $m_h$,
values different from the previous ones.
For instance we could assign to $m_{\tilde f}$ the smallest possible value
consistent with the LEP lower bound $m_{\tilde f} > 45~{\rm GeV}$[7] and
with the condition that the neutralino in the lightest SuSy particle, and take
for instance $m_{\tilde f} = 1.2~m_\c$, when $m_\c>45~{\rm GeV}$ and
$m_{\tilde f} = 45~{\rm GeV}$ otherwise. In this case lighter
sfermions make the $\c-\c$ annihilation for gaugino
configurations much more efficient than before. As a consequence, now
the fluxes for gamma--radiation and for radio--emission, which are
proportional to $<\sigma v>_0 / (<\sigma v>_{\rm int})^2$
 are smaller than in the previous case (where sfermions
were assumed to be very massive). However, our
calculations indicate that a situation similar to that depicted in Fig.4
still persists, i.e. in a $F_\gamma$ vs $\Omega h^2$ plot many neutralino
configurations densely populate a region where  $\Omega \sim  0.01 - 0.1$ with
values of $F_\gamma$ only very slightly below the experimental upper
bound.

    As a conclusion, we claim that the density distribution (1) of DM in our
Galaxy results in a pointlike source of high--energy ($E_\gamma > 100$ MeV)
gamma--radiation with fluxes of order $\sim 10^{-7}-10^{-8}~{\rm cm}^{-2}~{\rm
s}^{-1}$, detectable with the help of present and futures techniques.

\bigskip
\centerline{Acknowledgments}
\bigskip

We thank Vittorio de Alfaro for many useful discussions.
This work was supported in part by Research Funds of the Ministero
dell'Universit\`a e della Ricerca Scientifica e Tecnologica.

\vfill
\eject

{\bf References.}

\item{[1]} A.V. Gurevich and K.P. Zybin, Sov. Phys. JETP 67 (1988) 1
\item{[2]} V.S. Berezinsky, A.V. Gurevich and K.P. Zybin, Phys. Lett. B294
(1992)
221
\item{[3]} M. Davis, F.J. Summers and D. Schleger, Nature 359 (1992) 393;
A.N. Taylor and M. Rowan--Robinson, Nature 359 (1992) 396
\item{[4]} C.Alcock et al., Nature 365 (1993) 621; E.Aubourg et al.,
Nature 365 (1993) 623
\item{[5]} Y. B. Zeldovich, Astrophysika 6 (1970) 319
\item{[6]} A.Bottino, V.de Alfaro, N.Fornengo, G.Mignola and M.Pignone,
Astroparticle Physics 2 (1994) 67
\item{[7]} D.Decamp et al. (ALEPH Coll.), Phys. Rep. 216(1992)253.
\item{[8]} T. Sj\"ostrand, CERN-TH.6488/92.
\item{[9]} PDG, Phys. Rev. D45, Part II, June 1992
\item{[10]} Gamma--ray spectra with similar features were also considered by H.
U. Bengtsson, P.Salati and J. Silk, Nucl. Phys. B346 (1990) 129.
\item{[11]} L.Blitz, H. Bloemen, W. Hermsen and T.M. Bania, Astron. Astrophys.,
143 (1985) 267; J.Silk and H. Bloemen, Ap.J. Lett., 313 (1987) 4.
\item{[12]} N. Kardashev, Sov. Sci. Rev. Astroph. Space Phys. 4 (1985) 287
\item{[13]} W. Kundt, Astroph. and Space Sci., 172 (1990) 109
\item{[14]} M.G. Watson et al., Ap.J. 250 (1981) 142

\vfill
\eject

{\bf Figure Captions}

\bn

{\bf Figure 1:}  Scatter plot of the neutralino
relic abundance as a function of the neutralino mass. This scatter plot
has been obtained by varying $M_2$ and
$\mu$ in the ranges $0< M_2 < 6$ TeV and $-3 ~{\rm TeV} < \mu < 3~ {\rm
TeV}$ respectively.
The other parameters are taken at
the following values: $\tan \beta = 8$, $m_h=50~{\rm GeV}$,
$m_{\tilde f} = \tilde m = 3~{\rm TeV}$ ($m_{\tilde f}$  denotes the mass of
the sfermions, taken here to be degenerate)

\medskip

{\bf Figure 2:} Scatter plot of the integral gamma--ray flux
$F_\gamma(> 300~{\rm MeV})$ as a function of the neutralino mass. The
ranges for $M_2$ and $\mu$ are the same as in Fig.1. Also the other
parameters are taken at the same values as in Fig.1.

\medskip

{\bf Figure 3:} $\mu-M_2$ diagram where the squares denote
the $\c$--compositions
which satisfy the two following requirements: a) $\Omega_\c$ is
within the range: $0.5 \leq \Omega_\c \leq 1$ which,
taking $h=0.7$, implies $0.25 \leq \Omega_\c h^2 \leq 0.5$; b) the relevant
radio flux, as evaluated from eq.(9), does not exceed the experimental
upper bound $F_\nu <~ 1 \cdot 10^{-23}~{\rm erg}~{\rm cm}^{-2}~{\rm s}^{-1}~
{\rm Hz}$ at $\nu=1~{\rm GHz}$ [12,13]. The gamma-ray flux for these
configurations is
$ 5\cdot 10^{-9}~{\rm cm}^{-2}~{\rm s}^{-1}
\lsim F_{\gamma} \lsim 2\cdot 10^{-7} ~{\rm cm}^{-2}
{\rm s}^{-1}$

\medskip

{\bf Figure 4:} Fig. Scatter plot of $F_\gamma$ vs $\Omega_\c$ restricted
to those configurations with mass $20~{\rm GeV} \leq m_\c \leq 100~{\rm GeV}$,
which satisfy
the observational upper bound on radio--emission.

\medskip

\bye